\documentclass[aps,prd,nofootinbib,superscriptaddress,preprint,eqsecnum,showkeys,showpacs,preprintnumbers]{revtex4-2}
\usepackage{amsmath}
\usepackage{amssymb}
\usepackage{graphicx}
\usepackage{epsfig}
\usepackage{xcolor,color}
\usepackage{url}
\usepackage{bm}
\usepackage{mathrsfs}
\usepackage[utf8]{inputenc}
\usepackage{hyperref}
\usepackage{enumerate}
\usepackage{amsthm}
\usepackage{bbm}
\usepackage[normalem]{ulem}
\usepackage{upgreek}
\usepackage{tensor}
\usepackage{siunitx}
\usepackage{orcidlink}
\usepackage[caption=false]{subfig}
\usepackage{float}
\usepackage{colortbl}
\usepackage{hyperref} %引用等自动跳转
\usepackage{appendix} 
\usepackage{cleveref}
\usepackage{multirow}
\usepackage{booktabs}
\usepackage{threeparttable}
\usepackage[export]{adjustbox}
\usepackage{times}
\usepackage{commath}

\usepackage{orcidlink}

 \allowdisplaybreaks[3]
 
\hypersetup{colorlinks=true,linkcolor=blue,anchorcolor=blue,citecolor=blue}  %引用加蓝色
\definecolor{colour1}{HTML}{0571b0} %--- Blue
\definecolor{colour2}{HTML}{92c5de} %--- Cyan
\definecolor{colour3}{HTML}{f4a582} %--- Orange
\definecolor{colour4}{HTML}{ca0020} %--- Maroon
\definecolor{colour5}{HTML}{fe4a49} %--- Red
\definecolor{colour6}{HTML}{2d3092} %--- Red

\hypersetup{colorlinks=true, linkcolor=colour6, citecolor=colour6,
filecolor=colour6, urlcolor=colour6}

\theoremstyle{plain}

\DeclareMathOperator{\arccosh}{arccosh}

\newcommand{\bea}{\begin{eqnarray*}}
\newcommand{\eea}{\end{eqnarray*}}
\newcommand{\bean}{\begin{eqnarray}}
\newcommand{\eean}{\end{eqnarray}}

\newcommand{\vvr}{\mbox{\boldmath${r}$}}
\newcommand{\vp}{\mbox{\boldmath${p}$}}

\newcommand{\vS}{\mbox{\boldmath${S}$}}

\newcommand{\be}{\begin{equation}}
\newcommand{\ee}{\end{equation}}

\newcommand\beq{\begin{equation}}
\newcommand\eeq{\end{equation}}
\def\bea{\begin{eqnarray}}
\def\eea{\end{eqnarray}}

\graphicspath{{Figures/}}

\begin{document}
\title{Effective metric for binaries in framework of EOB theory to fifth PM order}

%\title{Effective metric for binaries in framework of effective one-body theory to fifth post-Minkowskian order}

\author{Jiliang Jing\orcidlink{0000-0002-2803-7900}\footnote{jljing@hunnu.edu.cn}}
 \affiliation{Department of Physics, Key Laboratory of Low Dimensional Quantum Structures and Quantum Control of Ministry of Education, and Synergetic Innovation
Center for Quantum Effects and Applications, Hunan Normal
University, Changsha, Hunan 410081, P. R. China}
\affiliation{Center for Gravitation and Cosmology, College of Physical Science and Technology, Yangzhou University, Yangzhou 225009, P. R. China}

\author{Weike Deng\orcidlink{0009-0002-5504-8151}
\footnote{wkdeng@hnit.edu.cn}} 
\affiliation{School of Science, Hunan Institute of Technology, Hengyang 421002, P. R. China}
\affiliation{Department of Physics, Key Laboratory of Low Dimensional Quantum Structures and Quantum Control of Ministry of Education, and Synergetic Innovation Center for Quantum Effects and Applications, Hunan Normal
University, Changsha, Hunan 410081, P. R. China}

\author{Sheng Long
\footnote{shenglong@ucas.ac.cn}}
\affiliation{School of Fundamental Physics and Mathematical Sciences, Hangzhou Institute for Advanced Study, University of Chinese Academy of Sciences, Hangzhou 310024, China}

%\date{\today}

\begin{abstract}
To establish a self-consistent effective one-body (EOB) theory that describes the dynamical evolution of binary systems based on the post-Minkowskian (PM) approximation—where the Hamiltonian, radiation reaction force, and waveforms are derived from an effective metric—the primary objective is to obtain the effective metric. Given that third-generation gravitational wave detectors require at least fifth-order PM accuracy, in this paper we constructed an effective metric in the EOB theory of  binaries up to fifth PM order. The effective metric is of type D, allowing for the derivation of decoupled and variable-separable equations for the null tetrad component of the gravitational perturbed Weyl tensor $ \psi^B_{4} $. This presents a basis for us to establish a self-consistent EOB theory up to 5PM order. 

\end{abstract}

\maketitle
\newpage

\section{Introduction}

Waveform templates play a crucial role in the detection of gravitational waves generated by binary systems ~\cite{Abbott2016,Abbott20162,Abbott2017,Abbott20172,Abbott20173,Abbott2019,Abbott20211,Abbott20212,Abbott20213}. The foundation for constructing these waveform templates is the theoretical model of gravitational radiation, with a key focus on the dynamical evolution during the late stages of binary mergers.

To investigate  the dynamical evolution for coalescing compact object binary systems, Buonanno and Damour \cite{Buonanno_1999} introduced an innovative approach that maps the two-body problem onto an EOB framework utilizing the post-Newtonian (PN) approximation. Subsequently, Damour et al. provided estimates of the gravitational waveforms emitted during the inspiral, plunge, and coalescence phases \cite{Buonanno_2000, Damour_2000}. This study was later generalized to incorporate spinning black holes \cite{Damour_2001, Buonanno_2006}. The EOB waveforms were further refined by calibrating the model against increasingly accurate numerical relativity simulations, thereby encompassing larger regions of the parameter space \cite{Buonanno_2007, Pan_2008,  Damour_2008, Damour_20082, Boyle_2008,Damour_20083, Pan_2010, Buonanno_2009, Damour_2009a, Pan_2011, Barausse_2012, Taracchini_2012, Damour_2015}. This refinement is crucial for the analysis of gravitational wave signals \cite{Taracchini_2014, Cao_2017}, and played a vital role in the analysis of the gravitational wave signals in the LIGO/Virgo collaboration.

In the EOB theory based on the PN approximation, the ratio $v/c$ 
 is typically considered a small quantity. To relax this constraint, Damour  \cite{Damour2016, Bini_2018}  proposed a novel theoretical model that integrates EOB theory with the PM approximation, as opposed to the PN approximation. Khalil et al.  \cite{Khalil_2022} demonstrated that the dynamics at the 4PM order align more closely with numerical relativity (NR) than those at the third 3PM order. Damour and Rettegno \cite{Damour_2023}  found that reformulating PM information in terms of EOB radial potentials yields remarkable agreement with NR data, particularly when utilizing radiation-reacted 4PM information. Recent significant advancements in PM theory have sparked considerable interest in developing waveform models that effectively incorporate information from various perturbative approaches. This integration is especially critical for addressing the accuracy challenges posed by upcoming detectors, including the Einstein Telescope, Cosmic Explorer, and space-based detectors such as LISA, TianQin, and Taiji.
 
We aim to develop a self-consistent EOB theory based on the PM approximation, wherein the Hamiltonian, radiation-reaction force, and waveform are all formulated within a unified physical model. In an EOB framework, the dynamical evolution of a coalescing binary system is governed by the Hamilton equations  \cite{Damour2001,Taracchini1}
\begin{subequations}\label{HEq}
  \begin{eqnarray}
  &&  \frac{d\vvr}{d\hat{t}}=\{\vvr,\hat{H}[g^{\text{eff}}_{\mu\nu}]\}=\frac{\partial \hat{H}[g^{\text{eff}}_{\mu\nu}]}{\partial \vp}\,,\\\label{EOM2}
 &&   \frac{d\vp}{d\hat{t}}=\{\vp,\hat{H}[g^{\text{eff}}_{\mu\nu}]\}+\hat{\bm{\mathcal{F}}}[g^{\text{eff}}_{\mu\nu}]=-\frac{\partial \hat{H}[g^{\text{eff}}_{\mu\nu}]}{\partial \vvr}
    +\hat{\bm{\mathcal{F}}}[g^{\text{eff}}_{\mu\nu}]\,,\\\label{EOM3}
&&   \frac{d\vS_{1,2}}{d\hat{t}} = \{\vS_{1,2}, \nu \hat{H}[g^{\text{eff}}_{\mu\nu}] \} = \nu \frac{\partial \hat{H}[g^{\text{eff}}_{\mu\nu}]}{\partial \vS_{1,2}} \times \vS_{1,2}  +\hat{\bm{\mathcal{F}}_{1,2}^s}\,, \label{EOM4}
    \end{eqnarray}
\end{subequations}
where $\hat{t}\equiv t/ M $,  $\nu=\frac{m_0}{ M }=\frac{m_1 \, m_2}{(m_1+m_2)^2}$,
$\hat{H}[g^{\text{eff}}_{\mu\nu}]=h_[g^{\text{eff}}_{\mu\nu}]/m_0$ is the reduced EOB Hamiltonian \cite{BarausseH,BarausseH1,Barausse},   $\hat{\bm{\mathcal{F}}}[g^{\text{eff}}_{\mu\nu}]=\bm{\mathcal{F}}[g^{\text{eff}}_{\mu\nu}]/m_0$ is the reduced RRF, and $\hat{\bm{\mathcal{F}}_{1,2}^s}=\bm{\mathcal{F}_{1,2}^s}/m_0$ is spin the reduced RRF  which can be written as \cite{PhysRevD.96.084064,Liu}
\begin{widetext}
\begin{align}
&\bm{\mathcal{F}_{1}^s}=\frac{4 p_r}{15r^4}(\pmb{L}\times\pmb{S}_1)\Big[\nu\Big(-\frac{22}{r}+36p^2-60p_r^2\Big)+\frac{1}{2}(1-2\nu-\delta)\Big(\frac{16}{r}-48p^2+75p_r^2\Big)\Big] \nonumber\\
&+\frac{2\nu}{15r^6}\Big[6r^2p_r(\pmb{S}_1\times\pmb{S}_2)\left(11\frac{1}{r}-18p^2+30p_r^2\right)-3r(\pmb{S}_1\times\pmb{p})\left(\pmb{r}\cdot\pmb{S}_2-\frac{1-\delta-2\nu}{2\nu}\pmb{r}\cdot\pmb{S}_1\right)  \nonumber\\
& \times \left(8\frac{1}{r}-9p^2+45p_r^2\right)  - (\pmb{r}\times\pmb{S}_1)\Big[15p_r\left(\pmb{r}\cdot\pmb{S}_2-\frac{1-\delta-2\nu}{2\nu}\pmb{r}\cdot\pmb{S}_1\right)\left(2\frac{1}{r}-3p^2+7p_r^2\right)\nonumber\\
&+2r\left(\pmb{p}\cdot\pmb{S}_2-\frac{1-\delta-2\nu}{2\nu}\pmb{p}\cdot\pmb{S}_1\right)\left(\frac{8}{r}-9p^2+45p_r^2\right)\Big]\Big]  - \frac{2(1-\delta)}{5r^5}\Big[4\left(\frac{1}{r}-3p^2+15p_r^2\right)\nonumber\\
&\times \Big((\pmb{r}\times\pmb{S}_1)+(\pmb{p}\times\pmb{S}_1)\Big)\pmb{r}\cdot\pmb{S}_1+\frac{15}{r}p_r(\pmb{r}\times\pmb{S}_1)(\pmb{r}\cdot\pmb{S}_1)(3p^2-7p_r^2)-30rp_r(\pmb{p}\times\pmb{S}_1)(\pmb{p}\cdot\pmb{S}_1)\Big],
\end{align}
\end{widetext}
where $\pmb{L}=\pmb{r}\times\pmb{p}$ is the orbital angular momentum, $p_r=\pmb{r}\cdot\pmb{p}/r$,  $\delta=(m_1-m_2)/M$, and $\bm{\mathcal{F}_{2}^s}=\bm{\mathcal{F}_{1}^s}(1\leftrightarrow 2)$.

Conversely, to derive the expression for the radiation-reaction force $ \dot{R}\,{\cal F}_{R}[g_{\mu\nu}^{\text{eff}}] + \dot{\varphi}\,{\cal F}_{\varphi}[g_{\mu\nu}^{\text{eff}}]=\frac{dE}{dt} $, where $ \frac{dE}{dt} = \frac{1}{4\pi G \omega^2}\int |\psi^B_{4}|^2 r^2 d\Omega $ \cite{Ref:poisson, TagoshiSasaki745}, and the waveform $ \frac{1}{2}(\ddot h_{+}-i\ddot h_{\times})=\psi^B_{4} $ of the ``plus" and ``cross" polarization modes of gravitational waves using standard general relativistic methods, we must derive the decoupled and variable-separable equation for the null tetrad component of the gravitational  perturbed Weyl tensor $ \psi^B_{4} $ in this effective spacetime. To achieve this objective, the effective metric must be of type D. We \cite{Jing1,Jing,JingXZ,JIng3} have identified the 3PM and 4PM effective metrics and derived the decoupled equations for $ \psi^{B}_{4} $ by employing different gauges. However, these results can only be extended to the slowly rotating background spacetime \cite{JingXZ}. Recently, we \cite{Jing2025} developed a self-consistent EOB theory for binary systems. This theory adopts a novel gauge that enables the separation of variables in the decoupled equations, applicable to general cases, including those involving rapid rotation. Noteworthy, this self-consistent EOB framework can be applied to any PM order.

It is well established that higher-order approximations yield more accurate results. Recently, Driesse et al. \cite{driesse2024conservativeblackholescattering} identified conservative black hole scattering at the fifth PM and first self-force order. Utilizing the worldline quantum field theory formalism, Driesse et al. \cite{Driesse_2025} computed the radiation-reacted impulse, scattering angle, radiated energy, and recoil of a classical black hole (or neutron star) scattering event at the fifth PM and sub-leading self-force orders (5PM-1SF). At orders $G^5$, Bini et al. \cite{Bini_2021} derived explicit theoretical expressions for the last two previously undetermined parameters describing the fifth post-Newtonian dynamics. Given that the third generation of gravitational wave detectors requires at least fifth PM order precision \cite{pürrer2019readyliesahead}, this paper aims to construct an effective metric up to fifth PM order for the effective EOB theory.

The remainder of the paper is organized as follows: In Sec. II, we present the scattering angles for both the real two-body and EOB systems up to the 5PM order. In Sec. III, we investigate the mapping relationship between the real relativistic energy $\mathcal{E}$ of the real two-body system and the effective relativistic energy $\mathcal{E}_0$ of the EOB system, and we obtain the effective rotating metric up to the 5PM order. Finally, conclusions and discussions are presented in the last section.

%\vspace{0.2cm}

\section{Scattering angles for real two-body and EOB systems}

The foundation of EOB theory lies in the identification of effective metrics, which can be determined by analyzing the scattering angles of both EOB and real two-body systems in a systematic, order-by-order approach. In this section, we will present the scattering angles for both the real two-body and EOB systems up to the 5PM order. 

\subsection{5PM order scattering angle  for real two-body system}

For a massive spinless binary system the Hamiltonian can be taken as  \cite{Bern_2019,Bern20192} 
\begin{eqnarray}\label{HR}
H(\vec{p},\vec{r})=\sqrt{|\vec{p}|^{\;2}+m_1^2}
+\sqrt{|\vec{p}|^{\;2}+m_2^2}+\sum_{i=1}^{\infty}c_i
\big{(}\frac{G}{|\vec{r}|}\big{)}^i,
\end{eqnarray}
where $m_1$ and $m_2$ are the masses of the two particles, $|\vec{p}|^2$ denotes $\vec{p}^{\;2}$,  $\vec{r}$ is the distance vector between particles, and  the explicit expressions of $c_i=c_i (p_\infty^2)$ can be found in Refs. \cite{Bern_2019,Bern20192,Dlapa_2022}\footnote{We should note that   $c_i$ in the Ref.  \cite{Bern_2019} is equal to $c_i/i!$ in the Ref. \cite{Dlapa_2022}}. In the following we will take
\begin{align}
&m=m_1+m_2\;,\;\;\;\;\;\; \mu=\frac{m_1m_2}{(m_1+m_2)}\;,\ \ \ \ \ \nu=\frac{m_1m_2}{(m_1+m_2)^2}\;,\nonumber\\
&E_1=\sqrt{\vec{p}^{\;2}+m_1^2}\;,\
E_2=\sqrt{\vec{p}^{\;2}+m_2^2}\;, \ \ \  \; E=E_1+E_2.
\end{align}
In spherical coordinates $\{t, r,\theta,\phi\}$, we have
$\vec{p}^{\;2}=p_r^2+\frac{p_{\theta}^2}{r^2}+
\frac{p_{\phi}^2}{r^2\sin^2\theta},$  $
\vec{r}\cdot\vec{p}=r p_r,$ and  $
\vec{p}^{\;2}\vec{r}^{\;2}-(\vec{r}\cdot\vec{p})^2
=p^2_{\theta}+\frac{p_{\phi}^2}{\sin^2\theta}.$
Without loss of generality, we set $\theta=\frac{\pi}{2}$ to denote the plane in which two particles reside. Then, we can express the reduced action as
\begin{align}
 S=-\mathcal{E}t+J\phi+S_r(r,\mathcal{E},J)\;,
\end{align}
where the energy $\mathcal{E}=-p_t$ and the angular momentum $J=p_{\phi}$ are two conservative quantities.   Using the Hamilton-Jacobi equation
\begin{align}
\frac{\partial S}{\partial t}+H(q,\frac{\partial S}{\partial q})=0, 
\end{align}
we can obtain the radial momentum $p_r^2=(\frac{d S_r}{dr})^2$, which up to the 5PM order is given by
\begin{eqnarray}\label{prsquare}
p_r^2&=&\frac{P_0r^2-J^2}{r^2}+P_1\Big(\frac{G}{r}\Big) +P_2\Big(\frac{G}{r}\Big)^2+P_3\Big(\frac{G}{r}\Big)^3+P_4\Big(\frac{G}{r}\Big)^4+P_5\Big(\frac{G}{r}\Big)^5,
\end{eqnarray}
with
$
P_0=\Big(\frac{
\mu }{ \Gamma}\Big)^2(\gamma^2-1),$ $ P_{1}
=2 m \mu^2 \Big(\frac{2 \gamma^2-1}{\Gamma}\Big),$ $ 
P_2=\frac{3 m^2 \mu^2}{2}\Big(\frac{5 \gamma^2-1}{\Gamma}\Big), 
$ 
and the explicit form of $P_{3}$,  $P_{4}$, and $P_{5}$ can be driven out by means of results in Refs. \cite{2210.05541,Dlapa_2022a,Dlapa_2023,Bini_2021,driesse2024conservativeblackholescattering}, where  $\gamma = \frac{1}{2} \frac{\mathcal{E}^2-m_1^2-m_2^2}{m_1m_2} $,  $   \Gamma =  \frac{\mathcal{E}}{m_1+m_2}$. 

By using the  definition scattering angle 
\begin{align}
\chi^{\text{Nor}}=-\pi+2 J \int_{r_{\rm min}}^{\infty}\frac{d r}{r^2\sqrt{p_r^2}}\;, \label{chireal}
\end{align}
where the minimum distance $r_\text{min}$ is determined by $p_r(r_{min})=0$, we can obtain the scattering angle up to 5PM order without the radiation-reaction effect, which  is  described as \cite{2210.05541,Dlapa_2022a,Dlapa_2023,Bini_2021,driesse2024conservativeblackholescattering}
\begin{eqnarray}
\chi^{\text{Nor}}=\chi^{\text{Nor}}_1\frac{G}{J}+\chi^{\text{Nor}}_2\Big(\frac{G}{J}\Big)^2+\chi^{\text{Nor}}_3 \Big(\frac{G}{J}\Big)^3+\chi^{\text{Nor}}_4 \Big(\frac{G}{J}\Big)^4 +\chi^{\text{Nor}}_5 \Big(\frac{G}{J}\Big)^5\;,
\end{eqnarray}
with
\begin{eqnarray}
&&\chi^{\text{Nor}}_1 = 2 \,m_1\, m_2 \,\frac{2\gamma^2-1}{\sqrt{\gamma^2-1}}, \  \nonumber \\  &&\chi^{\text{Nor}}_2=\frac{3\,\pi \,m_1^2\, m_2^2}{4}\, \frac{5\gamma^2-1}{\Gamma},\ \nonumber \\
&&\chi^{\text{Nor}}_3= 2\, \, \frac{\mu\sqrt{ (\gamma^2-1)}}{\Gamma}\, P_{3} +\frac{2}{\pi}\chi^{\text{Nor}}_1 \, \chi^{\text{Nor}}_2-\frac{(\chi^{\text{Nor}}_1)^3}{12} , \nonumber \\
&& \chi^{\text{Nor}}_4=\frac{3 \pi}{4} \, \frac{\mu^2 (\gamma^2-1)}{\Gamma^2} P_{4}+\frac{3\pi}{8}\chi^{\text{Nor}}_1 \chi^{\text{Nor}}_3 +\frac{3}{2\pi}(\chi^{\text{Nor}}_2)^2-\frac{3}{4}(\chi^{\text{Nor}}_1)^2 \chi^{\text{Nor}}_2+\frac{\pi}{32}(\chi^{\text{Nor}}_1)^4,\nonumber \\ 
&&\chi^{\text{Nor}}_5=\frac{8}{3} \, \frac{\mu^3 (\gamma^2-1)^{3/2}}{\Gamma^3}\, P_{5}-\Big[\frac{9}{80}(\chi^{\text{Nor}}_1)^5 -\frac{3}{\pi}(\chi^{\text{Nor}}_1)^3 \chi^{\text{Nor}}_2 
 +\frac{3}{2}(\chi^{\text{Nor}}_1)^2 \chi^{\text{Nor}}_3
 \nonumber \\ 
 &&\ \ \ \ \ \ -\frac{4}{\pi}\chi^{\text{Nor}}_2 \chi^{\text{Nor}}_3 -\frac{4\chi^{\text{Nor}}_1 (4\pi \chi^{\text{Nor}}_4-9(\chi^{\text{Nor}}_2)^2)}{3\pi^2}\Big],
 \label{6pmchi}
\end{eqnarray}
where $P_i=m^i \mu^2 \Gamma^{i-2} \tilde{P}_{i0}$ (i=3,4,5) in which the quantities  $\tilde{P}_{i0}$, for the sake of clarity, are presented in Appendix \ref{Pii}.

On the other hand,  the radiation-reaction effect up to the 5PM order can be expressed as\footnote{ Although some coefficients $J_4$ and $E_4$ have not yet been determined exactly, for the sake of convenience in future use, we provide the complete effective metric here.} \cite{Damour,Bini_2021,2210.05541,Manohar_2022} 
%$$
%\chi_3^{rr} = -\frac{2 \nu}{\Gamma^2} \frac{\gamma(2\gamma^2-1)^{2}}{3 (\gamma^2-1)^{3/2}} \left[ \frac{(5\gamma^2-8)\sqrt{\gamma^2-1}}{\gamma} + 2(9-6\gamma^2) \text{arcsinh}\sqrt{\frac{\gamma-1}{2}} \right],
%$$
\begin{eqnarray}
\label{PMexpchirad}
\chi^{\rm rr}_3 &=&\frac{\nu}{4}\hat{\chi}^{\text{Nor}}_1 J_2 ,
\,, \nonumber\\
\chi^{\rm rr}_4 &=&\frac{\nu}{4}\Big( \hat{\chi}^{\text{Nor}}_1 J_3 +2 \hat{\chi}^{\text{Nor}}_2 J_2 - \Gamma E_3  \frac{d \hat{\chi}^{\text{Nor}}_1}{d\gamma} \Big)
\,,\nonumber\\
\chi^{\rm rr}_5 &=&  \frac{\nu}{4}\Bigg[\hat{\chi}^{\text{Nor}}_1 J_4+2 \hat{\chi}^{\text{Nor}}_2 J_3+3 \hat{\chi}^{\text{Nor}}_3 J_2 
- \Gamma  \Big( E_4 \frac{d \hat{\chi}^{\text{Nor}}_1}{d\gamma} + E_3  \frac{d \hat{\chi}^{\text{Nor}}_2}{d\gamma} \Big)\Bigg],  \label{J234}
\end{eqnarray}
where $\hat{\chi}^{\text{Nor}}_i$ (i=1,2,3), $J_j $ (j=2, 3, 4), and $E_k$ (k=3,4) are presented in Appendix \ref{Jii}.

It is noteworthy that the relationship between our results $ \chi^{\text{Nor}}_i $ and the $ \chi_j^{(i)} $ from Refs. \cite{K_lin_2020, K_lin_2020a, Dlapa_2022} is expressed as $ \chi^{\text{Nor}}_i = 2(m_1 m_2)^i \chi_j^{(i)} $. This is because the PM coefficients of the scattering angle are defined by $ \frac{\chi}{2} = \sum_{i=1} \frac{\chi_j^{(i)}}{j^i} $, where $ j = \frac{J}{G m_1 m_2} $ in Refs. \cite{K_lin_2020, K_lin_2020a, Dlapa_2022}. We retain the notation $ J $ instead of substituting it with $ j $ because we will establish the relationship between $ J $ and $ J_0 $, where $ J_0 $ represents the angular momentum of the effective system. Consequently, the coefficients of the total scattering angle, including radiation-reaction effects, up to the 5PM order for the real two-body system are 
\begin{eqnarray}
\chi^{rel}_1 &=& 2 \,m_1\, m_2 \,\frac{2\gamma^2-1}{\sqrt{\gamma^2-1}}, \  \nonumber \\  \chi^{rel}_2&=&\frac{3\,\pi \,m_1^2\, m_2^2}{4}\, \frac{5\gamma^2-1}{\Gamma},\ \nonumber \\
\chi^{rel}_3&=& 2\, m_1^3\, m_2^3\, \chi_3^{rr}+2 \frac{\mu \sqrt{\gamma^2-1}}{\Gamma}\, P_{3} +\frac{2}{\pi}\chi^{rel}_1 \, \chi^{rel}_2-\frac{(\chi^{rel}_1)^3}{12} , \nonumber \\
 \chi^{rel}_4&=&2 m_1^4 m_2^4\chi^{r r}_4-\frac{3\pi }{4}m_1^3 m_2^3\chi^{rel}_1 \chi^{rr}_3+\frac{3 \pi}{4}\,\frac{\mu^2  (\gamma^2-1)}{\Gamma^2} P_{4}+\frac{3\pi}{8}\chi^{rel}_1 \chi^{rel}_3 +\frac{3}{2\pi}(\chi^{rel}_2)^2\nonumber \\    &-&\frac{3}{4}(\chi^{rel}_1)^2 \chi^{rel}_2+\frac{\pi}{32}(\chi^{rel}_1)^4,
 \nonumber \\ 
\chi^{rel}_5&=& 2 m_1^5 m_2^5 \chi^{r r}_5+\frac{8}{3} \, \frac{\mu^3  (\gamma^2-1)^{\frac{3}{2}}}{\Gamma^3}\, P_{5}+ m_1^3 m_2^3 \Big(3(\chi^{rel}_1)^2
 -\frac{8}{\pi}\chi^{rel}_2\Big) \chi^{\text{rr}}_3-\frac{32 m_1^4 m_2^4}{3 \pi}\chi^{\text{rr}}_4\chi^{rel}_1\nonumber \\ 
 &-&\Big[\frac{9}{80}(\chi^{rel}_1)^5 -\frac{3}{\pi}(\chi^{rel}_1)^3 \chi^{rel}_2 
 +\Big(\frac{3}{2}(\chi^{rel}_1)^2
 -\frac{4}{\pi}\chi^{rel}_2\Big) \chi^{rel}_3 -\frac{4\chi^{rel}_1 (4\pi \chi^{rel}_4-9(\chi^{rel}_2)^2}{3\pi^2}\Big]. \nonumber \\ 
\label{2pmchi1}
\end{eqnarray}

\subsection{5PM order Scattering angle  for  EOB system}

 We attempt to establish a self-consistent EOB theory, where ``self-consistent" means that the Hamiltonian, radiation reaction force, and waveforms within the theory are all based on the same physical model. To determine the expressions for the radiation-reaction force and waveform for the ``plus" and ``cross" polarization modes of gravitational waves by using standard general relativistic methods, we must derive the decoupled and variable-separable equation for $\psi^B_{4}=\frac{1}{2}(\ddot h_{+}-i\ddot h_{\times})$ within this effective spacetime.  To achieve this goal, the effective metric must be of Petrov Type D.  Therefore, we take the metric $g_{\mu\nu}^{\rm eff}$ as
\begin{eqnarray}
ds_{\rm eff}^2=g_{\mu\nu}^{\rm eff} d x^\mu d x^\nu =A dt^2-\frac{1}{A}d r^2- r^2(d\theta^2+\sin^2\theta d\phi^2),\label{Mmetric}
\end{eqnarray}
with
\begin{eqnarray}\label{expansionB}
&& A=1+\sum_{i=1}^\infty a_i \Big(\frac{GM_0}{r}\Big)^i,
\end{eqnarray}
where  $M_0$ is the mass parameter of the black hole, and $a_i$ ($i=1, 2, 3, 4, 5$) are dimensionless parameters which will be found in the following.

In the effective spacetime, the effective Hamilton-Jacobi equation reads
\begin{eqnarray}
g^{\mu\nu}_{\rm eff}\frac{\partial S_{\rm eff}}{\partial x^{\mu}}
\frac{\partial S_{\rm eff}}{\partial x^{\nu}}+m_0^2=0\;.\label{374H}
\end{eqnarray}
Due to the effective spacetime  (\ref{Mmetric}) has spherical symmetry, $S_{\rm eff}$ can be taken as
\begin{eqnarray}
S_{\rm eff}=-\mathcal{E}_0t+J_0\Phi+S_r^{0}(r,\mathcal{E}_0,J_0)\;,\label{375H}
\end{eqnarray}
where $\mathcal{E}_0$ and $J_0$ are the effective energy and angular momentum, respectively.
Substituting Eq.~\eqref{375H} into Eq.~\eqref{374H}, we obtain
\begin{eqnarray}
-\frac{\mathcal{E}_0^2}{A}+A
\Big{(}\frac{dS_r^0(r,\mathcal{E}_0,J_0)}{d r}\Big{)}^2+
\frac{J_0^2}{r^2}+m_0^2=0\;.
\end{eqnarray}
Then, we have
\begin{eqnarray}
\frac{dS_r^0(r,\mathcal{E}_0,J_0)}{dr}=\sqrt{\mathcal{R}_0(r,\mathcal{E}_0,J_0)}\;,
\end{eqnarray}
with
\begin{eqnarray}
\mathcal{R}_0(R,\mathcal{E}_0,J_0)=
\frac{1}{A^2}\mathcal{E}_0^2-\frac{1}{A}
\bigg{[}m_0^2+\frac{J_0^2}{r^2}\bigg{]}\;.\label{effR0}
\end{eqnarray}

For the EOB system, the  scattering angle is defined as
\begin{eqnarray}\label{chii}
\chi^{\text{eff}}&=&-\pi-2  \int_{r_{\rm min}}^{\infty}\frac{\partial\sqrt{ \mathcal{R}_0 (r,\mathcal{E}_0,J_0)}}{\partial J_0} d r.
\end{eqnarray}
where $r_\text{min}$ is the minimum distance  determined by setting the vanishing of $\mathcal{R}_s (r,\mathcal{E}_0,J_0)$.
Then, using Eqs. (\ref{Rii}) and (\ref{chiiA}) we have
\begin{eqnarray}
\label{x} \chi^{\text{eff}}=\chi^{\text{eff}}_1\frac{G}{J_0}+\chi^{\text{eff}}_2 \Big(\frac{G}{J_0}\Big)^2+\chi^{\text{eff}}_3 \Big(\frac{G}{J_0}\Big)^3+\chi^{\text{eff}}_4 \Big(\frac{G}{J_0}\Big)^4+\chi^{\text{eff}}_5 \Big(\frac{G}{J_0}\Big)^5,
 \end{eqnarray}
 with
\begin{align}\label{2pmchii}
\chi^{\text{eff}}_1&=\frac{2 M_0 \left[2 \mathcal{E}_0^2-m_0^2 \right]}{\sqrt{\mathcal{E}_0^2-m_0^2}},  \nonumber \\
\chi^{\text{eff}}_2&=\frac{M_0^2 \pi}{4}\bigg{[}\mathcal{E}_0^2(15 - 3 a_2 )+m_0^2( a_2-3)\bigg{]}, \nonumber  \\
\chi^{\text{eff}}_3&=\frac{2\chi^{\text{\text{\text{eff}}}}_1 \chi^{\text{eff}}_2}{\pi}-\frac{\chi^{\text{eff\,3}}_1}{12} -\frac{M_0^3 \sqrt{\mathcal{E}_0^2-m_0^2} }{3}
[(3 - 3 a_2 - 2 a_3)     m_0^2 + 2 \mathcal{E}_0^2( 15 a_2 + 4 a_3 -27)],  \nonumber\\
\chi^{\text{eff}}_4&=\frac{3\pi}{8}\chi^{\text{eff}}_1 \chi^{\text{eff}}_3 +\frac{3}{2\pi}(\chi^{\text{eff}}_2)^2-\frac{3}{4}(\chi^{\text{eff}}_1)^2 \chi^{\text{eff}}_2+\frac{\pi}{32}(\chi^{\text{eff}}_1)^4 + \frac{\pi M_0^4 (\mathcal{E}_0^2-m_0^2)}{64}\Big[(387-390 a_2  \nonumber \\
& +51 a_2^2  -164 a_3-60 a_4) \mathcal{E}_0^2-(3+3(a_2-2)a_2 - 4 a_3-12 a_4)m_0^2 \Big],
\nonumber \\ 
 \chi^{\text{eff}}_5&=\frac{2 M_0^5}{15} \left(\mathcal{E}_0^2-m_0^2\right)^{3/2} \Big[6 \left(20+10 a_{2}^2+6 a_{2} (a_{3}-5)-15 a_{3}-7 a_{4}-4
a_{5}\right) \mathcal{E}_0^2\nonumber \\ &- (a_{2} a_{3}+3 a_{4}-4 a_{5}) m_0^2\Big] 
-\Big[\frac{9}{80}(\chi^{\text{eff}}_1)^5 -\frac{3}{\pi}(\chi^{\text{eff}}_1)^3 \chi^{\text{eff}}_2 
+\frac{3}{2}(\chi^{\text{eff}}_1)^2 \chi^{\text{eff}}_3  -\frac{4}{\pi}\chi^{\text{eff}}_2 \chi^{\text{eff}}_3 \nonumber \\ 
 & 
  -\frac{4\chi^{\text{eff}}_1 (4\pi \chi^{\text{eff}}_4-9(\chi^{\text{eff}}_2)^2)}{3\pi^2}\Big].    \end{align}

\section{Energy map and effective metric  for 5PM EOB theory}

The map of the two-body problem onto an EOB problem can be realized by identifying the scattering angles for the two systems order by order, i.e., by taking $\chi^{rel}_i=\chi^{\text{eff}}_i\;, $ for $i=1,~2,~3,~4,\cdot \cdot\cdot $.  The energy map and the effective metric up to 5PM order for EOB theory are found in the following.

\subsection{Energy map for EOB theory}

For 5PM order, we still take the relations presented by Buonanno and Damour \cite{Damour1999,Damour2016}
\begin{eqnarray}
&&m_0=\frac{m_1m_2}{m_1+m_2},\label{i1}\nonumber \\
&&M_0=m_1+m_2,\nonumber \\
&&J_0=J.\label{i2}
\end{eqnarray}
By requiring that the effective metric at the 1PM order coincides with the Schwarzschild metric, and  taking
\begin{eqnarray}
\chi^{rel}_1=\chi^{\text{eff}}_1\;,\label{chi1}
\end{eqnarray}
we find the energy map between the relativistic energy $\mathcal{E}$ of the real two-body system and the relativistic energy $\mathcal{E}_0$ of the EOB system
\begin{eqnarray}
\mathcal{E}_0&=\frac{\mathcal{E}^2-m_1^2-m_2^2}{2(m_1+m_2)}\label{energymap1}\;. 
\label{E0}
\end{eqnarray}
Comparing Eq. (\ref{E0}) with $\gamma = \frac{1}{2} \frac{\mathcal{E}^2-m_1^2-m_2^2}{m_1m_2} $,  we have $\gamma=\frac{\mathcal{E}_0}{m_0}$.

\subsection{Effective metric in 5PM EOB theory}

By using the Eqs. \eqref{2pmchi1} and \eqref{2pmchii}, and taking 
\begin{eqnarray}
\chi^{rel}_i=\chi^{\text{eff}}_i\;, \ \ \ \ \text{(i=2,\ 3,\  4,\ 5)},\label{chi2}
\end{eqnarray}
we find that the parameters $a_i$ in the effective metric (\ref{Mmetric}) are given by
\begin{eqnarray} \label{parameters1}
a_2&=&\frac{3(1- \, \Gamma)(1-5 \, \gamma^2)}{\Gamma\, (3\, \gamma^2-1 )}\, ,\nonumber \\
a_3&=&\frac{3}{2 (4 \gamma^2-1 )}\Bigg[\frac{3- 2\, \Gamma-3 (15 -8\, \Gamma ) \gamma^2+6 ( 25-16 \, \Gamma ) \gamma^4}{\Gamma\, (3\, \gamma^2-1 )}-2 \tilde{P}_{30}-\frac{2 \chi_3^{rr}}{\sqrt{1-\gamma^2}}\Bigg], \nonumber \\
a_4&=&\frac{4}{(\gamma^2-1)(5 \, \gamma^2-1)}\Bigg\{\frac{2(2 \gamma^2-1) \chi_3^{rr}}{\sqrt{\gamma^2-1}} -\frac{8 \chi_4^{rr}}{3 \pi}+\frac{1}{48}\Big[ (\gamma^2-1)\Big( 387 \gamma^2-48 \tilde{P}_{40}-3\nonumber \\  &
  +&6 a_2 (1-65  \gamma^2) +3 a_2^2 (17  \gamma^2-1)+4 a_3 (1-41  \gamma^2)\Big) \Big]\Bigg\}, \nonumber \\     
  a_5&=& \frac{a_{2} a_{3}+3 a_{4}+20  \tilde{P}_{50}-6 \left(20+10 a_{2}^2+6 a_{2} (a_{3}-5)-15 a_{3}-7 a_{4}\right)
\gamma ^2}{4(1-6 \gamma ^2)}
  \nonumber \\   
 &+& \frac{1}{4 \pi  \left(\gamma ^2-1\right)^3
\left(6 \gamma ^2-1\right) \Gamma }\Bigg\{160 \left(1-3 \gamma ^2+2 \gamma ^4\right) \Gamma  \chi_4^{rr}+15 \pi  \sqrt{\gamma ^2-1} \Big[3 \Big(1
\nonumber \\ 
&-&\gamma ^2 (6-8 \Gamma )-2
\Gamma - \gamma ^4 (8 \Gamma-5 )\Big) \chi_3^{rr}-\left(\gamma ^2-1\right) \Gamma  \chi_5^{rr}\Big]\Bigg\}.    \end{eqnarray}

Based on the spinless effective metric, using the approach of constructing an effective rotating metric presented by Damour or Barausse et al. \cite{Barausse,Damour}, we find that the effective rotating metric for a real spin two-body system is described by
\begin{align}\label{effmetric}
 ds^{2}=g^{\text{eff}}_{\mu\nu}d x^\mu d x^\nu=\frac{\Delta_{r}-a^2\sin ^2\theta}{\Sigma}dt^{2}-\frac{\Sigma}{ \Delta_{r}} dr^{2}-\Sigma d\theta^{2}-\frac{\Lambda_{t} \sin^2\theta }{\Sigma} d\phi^{2}+\frac{2\omega_{j} \sin^2\theta}{ \Sigma}dt d\varphi,
\end{align}
with
\begin{align}
\nonumber & \Sigma=\overline{\rho} \overline{\rho}^{*}, \quad\overline{\rho}=r+i a \cos\theta, \quad \overline{\rho}^{*}=r-i a \cos\theta, \quad \Delta_{r}=\Delta^0_{r}+a^2, \\
 &\Lambda_{t}=\varpi^{4}-a^{2} \Delta_{r} \sin^{2}\theta, \quad\varpi=(r^{2}+a^{2})^{\frac{1}{2}},\quad\omega_{j}=a(a^{2}+r^{2}-\Delta_{r}),
\end{align}
where $a=\left|\vS_{metric}\right|/M $ is the rotational parameter where $\vS_{metric}=\vS_{metric}(\vS_1,\vS_2)$, and $ \Delta^0_{r}=r^2- 2 GM_0 r+\sum_{i=2}^\infty a_i \frac{(GM_0)^i}{r^{i-2}}$.  It is interesting to note that, starting from the spinless metric (\ref{Mmetric}), both the approaches presented by Damour or Barausse et al. will give the same effective rotating metric (\ref{effmetric}).

We can demonstrate that the effective spacetime described by Eq. (\ref{effmetric}) satisfies the conditions of Petrov type D, and that the null tetrad components of the trace-free Ricci tensor fulfill the conditions $\phi_{00} = \phi_{22} = 0$. Consequently, we can derive the decoupled  and variable-separable  equations for the null tetrad component of the gravitational perturbed Weyl tensor, $\Psi_4^B$, as referenced in \cite{Jing, Jing1,Jing2025}. This allows us to establish a self-consistent EOB theory in which the Hamiltonian, radiation-reaction forces, and waveforms for the ``plus" and ``cross" modes are all grounded in the same effective spacetime.

\section{Conclusions and discussions}\label{sec5}

The theoretical model of gravitational radiation provides the foundation for constructing waveform templates of gravitational waves produced by binary systems, particularly focusing on the dynamical evolution during the late stages of binary mergers. Our objective is to develop a self-consistent EOB theory for the dynamical evolution of binary systems, utilizing the PM approximation. Within this framework, the Hamiltonian, radiation-reaction force, and waveform are all formulated in terms of an effective metric, which underscores that the core of the EOB theory lies in deriving this effective metric.

To derive the radiation-reaction force and the waveforms of the ``plus" and ``cross" polarization modes of gravitational waves using standard general relativistic methods, it is essential to formulate the decoupled and variable-separable equation for the null tetrad component of the gravitational perturbed Weyl tensor $ \psi^B_{4} $ within this effective spacetime. To achieve this goal, the effective metric   must be satisfied the conditions of Petrov type D for any PM order. 

Higher-order approximations often provide more accurate results. Considering that third-generation gravitational wave detectors require at least fifth-order PM accuracy, we have successfully constructed an effective metric in the EOB theory of binary that includes radiation reaction effects  to 5PM order. This was achieved by systematically comparing the scattering angles of effective one-body systems with those of real binary systems, based on the fifth-order PM and first-order self-force conservative black hole scattering results from Driesse et al., as well as the 5PM scattering angles calculated by Bini et al.  The effective metric is of Petrov type D, allowing for the derivation of decoupled and variable-separable equations for the null tetrad component of the perturbed Weyl tensor $ \psi^B_{4} $. This presents a basis for us to establish a self-consistent EOB theory up to 5PM order.

\vspace{0.3cm}
\acknowledgments
%|--------------------------------------------------------------------|
{ We would like to thank professors Songbai Chen,  Qiyuan Pan and Xiaokai He for useful discussions on the manuscript. This work was supported by the Grant of NSFC No. 12035005, and National Key Research and Development  Program of China No. 2020YFC2201400.}  %|--------------------------------------------------------------------|

\newpage

\appendix
%\begin{appendices}

\section*{Appendix}

\section{The quantities \texorpdfstring{  $\tilde{P}_{i0}$}{}  in  \texorpdfstring{  $P_i=m^i \mu^2 \Gamma^{i-2} \tilde{P}_{i0}$}{} (i=3, 4, 5) }
 \label{Pii}

The quantities $\tilde{P}_{i0}$  are given by 

  \begin{eqnarray}\tilde{P}_{30} &=& \frac{18\gamma^2-1}{2\,\Gamma^2}+\frac{8\,\nu\, (3+12\gamma^2-4\gamma^4)}{\Gamma^2\, \sqrt{1-\gamma^2}} \mbox{arcsinh}\sqrt{\frac{1-\gamma}{2}}+\frac{\nu}{\Gamma^2}\Big(1-\frac{103}{3}\gamma-18 \gamma^2-\frac{2}{3} \gamma^3\nonumber  \\ & +& \frac{3 \,\Gamma\,(1-2\gamma^2)(1-5 \gamma^2)}{(1+\Gamma)(1+\gamma)}\Big),  \nonumber \\ 
  \tilde{P}_{40}&=& 
   \frac{4}{3 \pi (\gamma^2-1)}\hat{\chi}^{\text{Nor}}_4-\frac{1}{\gamma^2-1}\Big[\frac{1}{2}\hat{\chi}^{\text{Nor}}_1 \hat{\chi}^{\text{Nor}}_3 +\frac{2}{\pi^2}(\hat{\chi}^{\text{Nor}}_2)^2-\frac{1}{\pi}(\hat{\chi}^{\text{Nor}}_1)^2 \hat{\chi}^{\text{Nor}}_2+\frac{1}{24}(\hat{\chi}^{\text{Nor}}_1)^4\Big], \nonumber \\ 
 \tilde{P}_{50}&=&\frac{3}{8(\gamma^2-1)^{3/2}} \Bigg\{\hat{\chi}^{\text{Nor}}_5 +\Big[\frac{9}{80}(\hat{\chi}^{\text{Nor}}_1)^5 -\frac{3}{\pi}(\hat{\chi}^{\text{Nor}}_1)^3 \hat{\chi}^{\text{Nor}}_2 
 +\frac{3}{2}(\hat{\chi}^{\text{Nor}}_1)^2 \hat{\chi}^{\text{Nor}}_3
 -\frac{4}{\pi}\hat{\chi}^{\text{Nor}}_2 \hat{\chi}^{\text{Nor}}_3\nonumber \\ 
 &-&\frac{4\hat{\chi}^{\text{Nor}}_1 (4\pi \hat{\chi}^{\text{Nor}}_4-9(\hat{\chi}^{\text{Nor}}_2)^2}{3\pi^2}\Big]\Bigg\}, \label{f4} 
\end{eqnarray} 
with 
    \begin{eqnarray}
\hat{\chi}^{\text{Nor}}_1 &=& 2  \,\frac{2\gamma^2-1}{\sqrt{\gamma^2-1}}, \  \nonumber \\  
\hat{\chi}^{\text{Nor}}_2&=&\frac{3\,\pi }{4}\, \frac{5\gamma^2-1}{\Gamma},\ \nonumber \\
\hat{\chi}^{\text{Nor}}_3&=& 2\, \, \sqrt{\gamma^2-1}\Bigg[ \frac{18\gamma^2-1}{2\,\Gamma^2}+\frac{8\,\nu\, (3+12\gamma^2-4\gamma^4)}{\Gamma^2\, \sqrt{\gamma^2-1}} \mbox{arcsinh}\sqrt{\frac{\gamma-1}{2}}+\frac{\nu}{\Gamma^2}\Big(1-\frac{103}{3}\gamma\nonumber \\  &-&18 \gamma^2-\frac{2}{3} \gamma^3+\frac{3 \,\Gamma\,(1-2\gamma^2)(1-5 \gamma^2)}{(1+\Gamma)(1+\gamma)}\Big)\Bigg]+\frac{2}{\pi}\hat{\chi}^{\text{Nor}}_1 \, \hat{\chi}^{\text{Nor}}_2-\frac{(\hat{\chi}^{\text{Nor}}_1)^3}{12}, 
\nonumber \\    
\hat{\chi}^{\text{Nor}}_4&=&  \frac{2\,\pi\,  (\gamma^2-1)^2}{ \Gamma^3}\Bigg\{\frac{3 h_{61}}{128 (\gamma ^2-1)^3}
    +\nu\bigg[
      -\frac{3 h_3 \mathrm{K}^2\big(\frac{\gamma -1}{\gamma +1}\big)}{32 \big(\gamma ^2-1\big)^2}
      +\frac{3 h_4 \mathrm{E}\big(\frac{\gamma -1}{\gamma +1}\big) \mathrm{K}\big(\frac{\gamma -1}{\gamma +1}\big)}{32 \big(\gamma ^2-1\big)^2}
      +\frac{\pi ^2 h_5}{16(1- \gamma ^2)}
     \nonumber \\& +&\frac{3 h_{27} \log ^2\left(\frac{\gamma +1}{2}\right)}{4(1- \gamma ^2)}
      -\frac{h_6 \log \left(\frac{\gamma -1}{2}\right)}{32 \left(\gamma ^2-1\right)^2}
      +\frac{3 h_{15} \log \left(\frac{\gamma -1}{2}\right) \log \left(\frac{\gamma +1}{2}\right)}{16 \left(\gamma ^2-1\right)}
      -\frac{h_{22} \log \left(\frac{\gamma +1}{2}\right)}{32 \left(\gamma ^2-1\right)^2} \nonumber \\&      -&\frac{h_{23} \log (\gamma )}{4 \left(\gamma ^2-1\right)^2}
      +\frac{3 h_{26} \arccosh^2(\gamma )}{64 \left(\gamma ^2-1\right)^4}
      +\frac{h_{24} \arccosh(\gamma )}{32 \left(\gamma ^2-1\right)^{7/2}}
      -\frac{3 h_{16} \log \left(\frac{\gamma -1}{2}\right) \arccosh(\gamma )}{32 \left(\gamma ^2-1\right)^{5/2}}
     \nonumber \\&    -&\frac{3 h_{28} \log \left(\frac{\gamma +1}{2}\right) \arccosh(\gamma )}{32 \left(\gamma ^2-1\right)^{5/2}}  -\frac{h_{62}}{384 \gamma ^7 \left(\gamma ^2-1\right)^3}
      -\frac{21 h_2 \mathrm{E}^2\left(\frac{\gamma -1}{\gamma +1}\right)}{64 (\gamma -1)^2 (\gamma +1)}
     \nonumber \\&   -&\frac{3 \sqrt{\gamma ^2-1} h_7 \text{Li}_2\left(\sqrt{\frac{\gamma -1}{\gamma +1}}\right)}{2 (\gamma -1)^2 (\gamma +1)^3}
      +\frac{h_{29} \text{Li}_2\left(\frac{1-\gamma }{\gamma +1}\right)}{8(1- \gamma^2)}  +\Big(\frac{3 \sqrt{\gamma ^2-1} h_7}{8 (\gamma -1)^2 (\gamma +1)^3}
      +\frac{3 h_{30}}{16-16 \gamma ^2}\Big)\nonumber \\
      & \times & \text{Li}_2\big(\frac{\gamma -1}{\gamma +1}\big)
      \bigg]\Bigg\},  \nonumber \\ 
\hat{\chi}^{\text{Nor}}_5&=&\sum_{k=1}^{31}
  c_k(\gamma) f_k(\gamma)\,, 
\label{2pmchi}
\end{eqnarray}
where 
 \begin{align}
            h_1 &= 515 \gamma ^6-1017 \gamma ^4+377 \gamma ^2-3,\nonumber \\
            h_2 &= 380 \gamma ^2+169,\nonumber\\
            h_3 &= 1200 \gamma ^2+2095 \gamma +834,\nonumber\\
            h_4 &= 1200 \gamma ^3+2660 \gamma ^2+2929 \gamma +1183,\nonumber\\
            h_5 &= -25 \gamma ^6+30 \gamma ^4+60 \gamma ^3-129 \gamma ^2+76 \gamma -12,\nonumber\\
            h_6 &= 210 \gamma ^6-552 \gamma ^5+339 \gamma ^4-912 \gamma ^3+3148 \gamma ^2-3336 \gamma +1151,\nonumber\\
            h_7 &= -\gamma  \left(2 \gamma ^2-3\right) \left(15 \gamma ^2-15 \gamma +4\right),\nonumber\\
            h_8 &= 420 \gamma ^9+3456 \gamma ^8-1338 \gamma ^7-15822 \gamma ^6+13176 \gamma ^5+9563 \gamma ^4-16658 \gamma ^3,\nonumber\\
            &\quad+8700 \gamma ^2-496 \gamma -1049,\nonumber\\
            h_9 &= -22680 \gamma ^{21}+11340 \gamma ^{20}+116100 \gamma ^{19}-34080 \gamma ^{18}-216185 \gamma ^{17}+74431 \gamma ^{16},\nonumber\\
            &\quad+232751 \gamma ^{15}-304761 \gamma ^{14}+333545 \gamma ^{13}-32675 \gamma ^{12}-500785 \gamma ^{11}+535259 \gamma ^{10},\nonumber\\
            &\quad-181493 \gamma ^9+3259 \gamma ^8+9593 \gamma ^7+9593 \gamma ^6-3457 \gamma ^5-3457 \gamma ^4,\nonumber\\
            &\quad+885 \gamma ^3+885 \gamma ^2-210 \gamma -210,\nonumber\\
            h_{10} &= -280 \gamma ^7+50 \gamma ^6+970 \gamma ^5+27 \gamma ^4-1432 \gamma ^3+444 \gamma ^2+366 \gamma -129,\nonumber\\
            h_{11} &= 2835 \gamma ^{11}-10065 \gamma ^9-700 \gamma ^8+13198 \gamma ^7+1818 \gamma ^6-9826 \gamma ^5+5242 \gamma ^4,\nonumber\\
            &\quad+11391 \gamma ^3+18958 \gamma ^2+10643 \gamma +2074,\nonumber\\
            h_{12} &= \gamma  \left(945 \gamma ^{10}-2955 \gamma ^8+4874 \gamma ^6-5014 \gamma ^4+8077 \gamma ^2+5369\right),\nonumber\\
            h_{13} &= \gamma  \left(280 \gamma ^7+580 \gamma ^6+90 \gamma ^5-856 \gamma ^4-2211 \gamma ^3+1289 \gamma ^2+2169 \gamma -1965\right),\nonumber\\
            h_{14} &= \gamma  \left(2 \gamma ^2-3\right) \left(280 \gamma ^7-890 \gamma ^6-610 \gamma ^5+1537 \gamma ^4+380 \gamma ^3-716 \gamma ^2-82 \gamma +85\right),\nonumber\\
            h_{15} &= 35 \gamma ^4+60 \gamma ^3-150 \gamma ^2+76 \gamma -5,\nonumber\\
            h_{16} &= \gamma  \left(2 \gamma ^2-3\right) \left(35 \gamma ^4-30 \gamma ^2+11\right),\nonumber \\
            h_{17} &= 315 \gamma ^8-860 \gamma ^6+690 \gamma ^4-960 \gamma ^3+1732 \gamma ^2-1216 \gamma +299,\nonumber \\
            h_{18} &= 315 \gamma ^6-145 \gamma ^4+65 \gamma ^2+21,\nonumber \\
            h_{19} &= 840 \gamma ^9+1932 \gamma ^8+234 \gamma ^7-17562 \gamma ^6+20405 \gamma ^5-2154 \gamma ^4-11744 \gamma ^3,\nonumber \\
            &\quad+12882 \gamma ^2-4983 \gamma +102,\nonumber \\
            h_{20} &= 3600 \gamma ^{16}+4320 \gamma ^{15}-23840 \gamma ^{14}+7824 \gamma ^{13}+14128 \gamma ^{12}+16138 \gamma ^{11}-9872 \gamma ^{10},\nonumber \\
            &\quad-47540 \gamma ^9+63848 \gamma ^8-37478 \gamma ^7+13349 \gamma ^6-1471 \gamma ^4+207 \gamma ^2-45,\nonumber \\
            h_{21} &= -350 \gamma ^7+1425 \gamma ^5-400 \gamma ^4-1480 \gamma ^3+660 \gamma ^2+285 \gamma -124,\nonumber \\
            h_{22} &= -300 \gamma ^7+210 \gamma ^6+1112 \gamma ^5+2787 \gamma ^4+2044 \gamma ^3+3692 \gamma ^2+6744 \gamma +1759,\nonumber \\
            h_{23} &= \gamma  \left(75 \gamma ^6-140 \gamma ^4-283 \gamma ^2-852\right),\nonumber \\
            h_{24} &= \gamma  \left(2 \gamma ^2-3\right) \left(210 \gamma ^6-720 \gamma ^5+339 \gamma ^4-576 \gamma ^3+3148 \gamma ^2-3504 \gamma +1151\right),\nonumber \\
            h_{25} &= \gamma  \left(2 \gamma ^2-3\right) \left(350 \gamma ^7-960 \gamma ^6-705 \gamma ^5+1632 \gamma ^4+432 \gamma ^3-768 \gamma ^2-93 \gamma +96\right),\nonumber \\
            h_{26} &= \gamma ^2 \left(3-2 \gamma ^2\right)^2 \left(35 \gamma ^4-30 \gamma ^2+11\right),\nonumber \\
            h_{27} &= 15 \gamma ^3+60 \gamma ^2+19 \gamma +8,\nonumber \\
            h_{28} &= \gamma  \left(70 \gamma ^6-645 \gamma ^4+768 \gamma ^2+63\right),\nonumber \\
            h_{29} &= -75 \gamma ^6+90 \gamma ^4+333 \gamma ^2+60
       ,\nonumber \\ 
            h_{30} &= 25 \gamma ^6-30 \gamma ^4+60 \gamma ^3+129 \gamma ^2+76 \gamma +12,\nonumber \\
\if
            h_{31} &= \left(1-5 \gamma ^2\right)^2,\nonumber \\
            h_{32} &= 80 \gamma ^8-192 \gamma ^6+152 \gamma ^4-44 \gamma ^2+3,\nonumber \\
            h_{33} &= \gamma  \left(2 \gamma ^2-1\right) \left(64 \gamma ^6-216 \gamma ^4+258 \gamma ^2-109\right),\nonumber \\
            h_{34} &= \left(2 \gamma ^2-1\right)^3 \left(5 \gamma ^2-8\right),\nonumber \\
            h_{35} &= \gamma  \left(2 \gamma ^2-3\right) \left(2 \gamma ^2-1\right)^3,\nonumber \\
            h_{36} &= 8 \gamma ^6-28 \gamma ^4+6 \gamma ^2+3,\nonumber \\
            h_{37} &= \gamma  \left(384 \gamma ^8-1528 \gamma ^6+384 \gamma ^4+2292 \gamma ^2-1535\right),\nonumber \\
            h_{38} &= 393897472 \gamma ^{16}-791542442 \gamma ^{14}-3429240286 \gamma ^{12}+3966858415 \gamma ^{10},\nonumber  \\
            &\quad+767410066 \gamma ^8-21241500 \gamma ^6+7188300 \gamma ^4-1837500 \gamma ^2+385875,\nonumber \\
            h_{39} &= 1575 \gamma ^7-2700 \gamma ^6-3195 \gamma ^5+3780 \gamma ^4+4993 \gamma ^3-1188 \gamma ^2-1485 \gamma +108,\nonumber \\
            h_{40} &= -3592192 \gamma ^{18}+2662204 \gamma ^{16}+46406238 \gamma ^{14}-37185456 \gamma ^{12}-25426269 \gamma ^{10},\nonumber \\
            &\quad+222810 \gamma ^8-246540 \gamma ^6+79800 \gamma ^4-19950 \gamma ^2+3675,\nonumber \\
            h_{41} &= 44 \gamma ^6-32 \gamma ^4-425 \gamma ^2-82,\nonumber \\
            h_{42} &= \gamma  \left(16 \gamma ^6+24 \gamma ^4-226 \gamma ^2-151\right),\nonumber \\
            h_{43} &= \gamma ^2 \left(4 \gamma ^8-59 \gamma ^4+35 \gamma ^2+60\right),\nonumber \\
            h_{44} &= -525 \gamma ^7+1065 \gamma ^5-3883 \gamma ^3+1263 \gamma,\nonumber \\
            h_{45} &= 175 \gamma ^7-150 \gamma ^6-355 \gamma ^5+210 \gamma ^4+185 \gamma ^3-66 \gamma ^2-37 \gamma +6,\nonumber \\
            h_{46} &= -175 \gamma ^7+355 \gamma ^5-185 \gamma ^3+37 \gamma,\nonumber \\
            h_{47} &= \gamma  \left(525 \gamma ^6-1065 \gamma ^4-2773 \gamma ^2+1041\right),\nonumber \\
            h_{48} &= 96 \gamma ^{10}-8464 \gamma ^8+54616 \gamma ^6-70104 \gamma ^4+9916 \gamma ^2+13895,\nonumber \\
            h_{49} &= 6144 \gamma ^{16}-587336 \gamma ^{14}+4034092 \gamma ^{12}-417302 \gamma ^{10}-5560073 \gamma ^8-142640 \gamma ^6,\nonumber \\
            &\quad+35710 \gamma ^4-8250 \gamma ^2+1575,\nonumber \\
            h_{50} &= -3747 \gamma ^6+3249 \gamma ^4+8535 \gamma ^2+1051,\nonumber \\
            h_{51} &= 24576 \gamma ^{18}+213480 \gamma ^{16}-1029342 \gamma ^{14}-1978290 \gamma ^{12}+3752006 \gamma ^{10}+816595 \gamma ^8 \nonumber\\
            &\quad-55260 \gamma ^6+13690 \gamma ^4-3100 \gamma ^2+525,\nonumber\\
            h_{52} &= \gamma  \left(16 \gamma ^6+204 \gamma ^4-496 \gamma ^2-869\right),\nonumber \\
            h_{53} &= \gamma ^2 \left(8 \gamma ^4-6 \gamma ^2-9\right),\nonumber \\
            h_{54} &= \gamma  \left(2 \gamma ^2-3\right) \left(8 \gamma ^6-6 \gamma ^4-51 \gamma ^2-8\right),\nonumber \\
            h_{55} &= -4321 \gamma ^6+3387 \gamma ^4+15261 \gamma ^2+2057,\nonumber \\
            h_{56} &= 2100 \gamma ^7-4996 \gamma ^6+1755 \gamma ^5+4332 \gamma ^4-6422 \gamma ^3+4212 \gamma ^2-1209 \gamma +36,\nonumber \\
            h_{57} &= -1249 \gamma ^6+1083 \gamma ^4+1053 \gamma ^2+9,\nonumber \\
            h_{58} &= -1823 \gamma ^6+1221 \gamma ^4+13155 \gamma ^2+2039,\nonumber \\
            h_{59} &= -24 \gamma ^6+18 \gamma ^4+111 \gamma ^2+16 ,\nonumber \\
            h_{60} &= \gamma  \left(26 \gamma ^2-9\right),\nonumber \\ 
        \fi       
                       h_{61} &= 35 (\gamma -1) (\gamma +1) \left(33 \gamma ^4-18 \gamma ^2+1\right), \nonumber \\
       h_{62} &= 3600 \gamma ^{16}+4320 \gamma ^{15}-35360 \gamma ^{14}+33249 \gamma ^{13}+27952 \gamma ^{12}-25145 \gamma ^{11}-15056 \gamma ^{10}\nonumber\\
       &-32177 \gamma ^9+64424 \gamma ^8-38135 \gamma ^7+13349 \gamma ^6-1471 \gamma ^4, +207 \gamma ^2-45, \nonumber \\
       h_{63} &= \gamma ^2 \left(2 \gamma ^2-3\right) \left(2 \gamma ^2-1\right) \left(35 \gamma ^4-30 \gamma ^2+11\right), \nonumber\\
       h_{64} &= -4140 \gamma ^8+702 \gamma ^7+15018 \gamma ^6-8491 \gamma ^5-9366 \gamma ^4+10052 \gamma ^3-6210 \gamma ^2+2681 \gamma -102, \nonumber\\
       h_{65} &= 210 \gamma ^7-240 \gamma ^6-755 \gamma ^5+216 \gamma ^4+1200 \gamma ^3-508 \gamma ^2-295 \gamma +124, \nonumber\\
       h_{66} &= \gamma  \left(2 \gamma ^2-3\right) \left(2 \gamma ^2-1\right) \left(35 \gamma ^4-30 \gamma ^2+11\right), \nonumber\\
       h_{67} &= -(\gamma -1) \Big(420 \gamma ^9+7596 \gamma ^8-2040 \gamma ^7-30840 \gamma ^6+21667 \gamma ^5+18929 \gamma ^4-26710 \gamma ^3\nonumber \\ &+14910 \gamma ^2-3177 \gamma -947\Big), \nonumber \\
       h_{68} &= (\gamma -1) \left(490 \gamma ^7-290 \gamma ^6-1725 \gamma ^5+189 \gamma ^4+2632 \gamma ^3-952 \gamma ^2-661 \gamma +253\right) ,
\end{align}  
 and the coefficient polynomials of the scattering angle $\hat{\chi}^{\text{Nor}}_5  
  = \sum_{k=1}^{31} c_k(\gamma) f_k(\gamma)$  are\begin{align}		
f_{1}(\gamma)&=1,\nonumber \\ 
f_{2}(\gamma)&=G(1;y),\nonumber \\ 
f_{3}(\gamma)&=2 (G(0;y)-G(1;y)+G(2;y)+\log (2)),\nonumber \\ 
f_{4}(\gamma)&=\pi ^2,\nonumber \\ 
f_{5}(\gamma)&=G(1;y)^2,\nonumber \\ 
f_{6}(\gamma)&=2 G(1;y) (G(0;y)-G(1;y)+G(2;y)+\log (2)),\nonumber \\ 
f_{7}(\gamma)&=\frac{1}{2} G(1;y)^2+(-G(1;y)+G(1-i;y)+G(1+i;y)) G(1;y)-G(1,1-i;y)\nonumber \\  & -G(1,1+i;y),\nonumber \\ 
f_{8}(\gamma)&=-\frac{1}{2} G(1;y)^2+G(2;y) G(1;y)+G(0,1;y)-G(1,2;y),\nonumber \\ 
f_{9}(\gamma)&=-G(1;y) G(2;y)+G(0,1;y)+G(1,2;y),\nonumber \\ 
f_{10}(\gamma)&=\pi ^2 G(1;y),\nonumber \\ 
f_{11}(\gamma)&=-G(1;y) \Big(-G(1;y)^2-2 G(0;y) G(1;y)+G(2;y) G(1;y)-\log (4) G(1;y)\nonumber \\ & -G(0,1;y)+4 G(1,1-i;y)+4 G(1,1+i;y)-G(1,2;y)\Big),\nonumber \\ 
f_{12}(\gamma)&=-G(1;y)^3+2 G(2;y) G(1;y)^2+(G(1;y) G(2;y)-G(0,1;y)-G(1,2;y)) G(1;y)\nonumber \\ &+4 \left(-\frac{1}{6} G(1;y)^3+G(1,1,1-i;y)+G(1,1,1+i;y)\right),\nonumber \\ 
f_{13}(\gamma)&=\frac{1}{4} G(1;y)^3-G(2;y) G(1;y)^2+G(2;y)^2 G(1;y)-G(0,1;y) G(1;y)\nonumber \\ &+G(1,2;y) G(1;y)+2 G(0,0,1;y)+G(0,1,1;y)-G(1,1,2;y)-2 G(1,2,2;y),\nonumber \\ 
f_{14}(\gamma)&=\left(-\frac{1}{2} G(1;y)^2+G(2;y) G(1;y)+G(0,1;y)-G(1,2;y)\right) (2 G(0;y) \nonumber \\ & -G(1;y)+\log (4)),\nonumber \\ 
f_{15}(\gamma)&=-\frac{1}{6} G(1;y)^3+G(1,1-i;y) G(1;y)+G(1,1+i;y) G(1;y)+(2 G(2;y)\nonumber \\ & -G(1;y)) \left(-\frac{1}{2} G(1;y)^2+G(1,1-i;y)+G(1,1+i;y)\right)-2 G(1,1,1-i;y)
\nonumber \\ & -2 G(1,1,1+i;y)+2 G(1,1,2;y)-2 G(1,1-i,2;y)-2 G(1,1+i,2;y),\nonumber \\ 
f_{16}(\gamma)&=(-G(1;y) G(2;y)+G(0,1;y)+G(1,2;y)) (2 G(0;y)-G(1;y)+\log (4)),\nonumber \\ 
f_{17}(\gamma)&=2 \left(-\frac{1}{6} G(1;y)^3+G(0,1,1;y)+G(1,1,2;y)\right)-\frac{3}{2} G(1;y) (-G(1;y) G(2;y)\nonumber \\ & +G(0,1;y)+G(1,2;y)),\nonumber \\ 
f_{18}(\gamma)&=\frac{1}{6} G(1;y)^3-(-G(1;y)+G(1-i;y)+G(1+i;y)) (2 G(2;y)-G(1;y)) G(1;y)\nonumber \\ & -G(1,1-i;y) G(1;y)-G(1,1+i;y) G(1;y)
-(G(1;y)-G(1-i;y)\nonumber \\ &-G(1+i;y)) (-2 G(0;y)+G(1;y)+\log (4)) G(1;y)-4 (-G(1;y)+G(1-i;y)\nonumber \\ & +G(1+i;y)) \log (2) G(1;y)
+2 G(1,1,1-i;y)+2 G(1,1,1+i;y)-2 G(1,1,2;y)\nonumber \\ & +2 G(1,1-i,2;y)+2 G(1,1+i,2;y)
+\Big(-\frac{1}{2} G(1;y)^2+G(1,1-i;y)\nonumber \\ & +G(1,1+i;y)\Big) (2 G(0;y)-G(1;y)-\log (4))+4 \Big(-\frac{1}{2} G(1;y)^2\nonumber \\ & +G(1,1-i;y)+G(1,1+i;y)\Big) \log (2),\nonumber \\ 
f_{19}(\gamma)&=G(1;y) \left(-\frac{1}{2} G(1;y)^2+G(2;y) G(1;y)+G(0,1;y)-G(1,2;y)\right),\nonumber \\ 
f_{20}(\gamma)&=(-G(1;y)+G(1-i;y)+G(1+i;y)) \left(2 G(0,1;y)-\frac{1}{2} G(1;y)^2\right)+(-G(1;y)\nonumber \\ & +G(1-i;y)+G(1+i;y)) \left(2 (G(1;y) G(2;y)-G(1,2;y))-\frac{1}{2} G(1;y)^2\right)
\nonumber \\ & -2 (-G(0,1,1;y)+G(0,1,1-i;y)+G(0,1,1+i;y))+2 (-G(1;y) G(1,2;y)\nonumber \\ & +2 G(1,1,2;y)+G(1,2,1-i;y)+G(1,2,1+i;y)),\nonumber \\ 
f_{21}(\gamma)&=2 G(1;y) (G(1;y) G(2;y)-G(0,1;y)-G(1,2;y)),\nonumber \\ 
f_{22}(\gamma)&=\frac{1}{2} \Bigl(\frac{1}{6} G(1;y)^3-G(1,1-i;y) G(1;y)-G(1,1+i;y) G(1;y)+\Bigl(\frac{1}{2} G(1;y)^2\nonumber \\ & -G(1,1-i;y)-G(1,1+i;y)+2 (-G(1;y) G(2;y)+G(1,2;y)+G(2,1-i;y)
\nonumber \\ & +G(2,1+i;y))\Bigr) G(1;y)+2 (-G(0,1,1;y)+G(0,1-i,1;y)+G(0,1+i,1;y))\nonumber \\ & +2 G(1,1,1-i;y)+2 G(1,1,1+i;y)\Bigr),\nonumber \\ 
f_{23}(\gamma)&=\frac{1}{12} G(1;y)^3-G(2;y) G(1;y)^2+G(2;y)^2 G(1;y)+G(0,1;y) G(1;y)\nonumber \\ & +G(1,2;y) G(1;y)-2 G(0,0,1;y)-G(0,1,1;y)-G(1,1,2;y)-2 G(1,2,2;y),\nonumber \\ 
f_{24}(\gamma)&=\frac{1}{2} G(1;y)^3-6 G(0,1,1;y)+4 G(0,1,2;y)+8 G(0,2,1;y)-4 (G(1;y) G(1,2;y)\nonumber \\ & -2 G(1,1,2;y))-2 G(1,1,2;y),\nonumber \\ 
f_{25}(\gamma)&=-\frac{1}{12} G(1;y)^3+G(0,1,1;y)-2 G(0,1,2;y)+G(1,1,2;y),\nonumber \\ 
f_{26}(\gamma)&=\frac{1}{2} \Bigl(-\frac{1}{6} G(1;y)^3+G(1,1-i;y) G(1;y)+G(1,1+i;y) G(1;y)+\Bigl(\frac{1}{2} G(1;y)^2\nonumber \\ & -G(1,1-i;y)-G(1,1+i;y)+2 (-G(1;y) G(2;y)+G(1,2;y)
\nonumber \\ &+G(2,1-i;y)+G(2,1+i;y))\Bigr) G(1;y)-2 (-G(0,1,1;y)+G(0,1-i,1;y)\nonumber \\ &+G(0,1+i,1;y))-2 G(1,1,1-i;y)-2 G(1,1,1+i;y)\Bigr),\nonumber \\ 
f_{27}(\gamma)&=\frac{1}{6} G(1;y)^3+\frac{1}{2} (-G(1;y) +G(1-i;y)+G(1+i;y))^2 G(1;y)\nonumber \\ & -G(1,1-i;y) G(1;y)-G(1,1+i;y) G(1;y)
-(-G(1;y)+G(1-i;y)\nonumber \\ & +G(1+i;y)) \left(-\frac{1}{2} G(1;y)^2+G(1,1-i;y)+G(1,1+i;y)\right)+G(1,1,1-i;y)\nonumber \\ & +G(1,1,1+i;y)
+G(1,1-i,1-i;y)+G(1,1-i,1+i;y)\nonumber \\ & +G(1,1+i,1-i;y)+G(1,1+i,1+i;y),\nonumber \\ 
f_{28}(\gamma)&=G(1;y)^3,\nonumber \\ 
f_{29}(\gamma)&=G(1;y)^2 \Bigl(-G(1;y)+G(1-i;y)+G(1+i;y)\Bigr),\nonumber \\ 
f_{30}(\gamma)&=-\Bigl(G(1;y)-2 G(2;y)\Bigr) \left(2 G(0,1;y)-\frac{1}{2} G(1;y)^2\right),\nonumber \\ 
f_{31}(\gamma)&= \Bigl(G(1;y)-2 G(2;y) \Bigr) \left(\frac{1}{2} G(1;y)^2-2 (G(1;y) G(2;y)-G(1,2;y))\right),
\end{align}
and 
\begin{align}
c_{1}(\gamma)&=\frac{1}{7560 (\gamma^{2} -1)^4 \gamma ^7 (\gamma +1)}\Big[1880064 \gamma ^{19}+1880064 \gamma ^{18}+42654086 \gamma ^{17}+20978054 \gamma ^{16}\nonumber \\ & -305752626 \gamma ^{15}-236079666 \gamma ^{14}+597683406 \gamma ^{13}+516398286 \gamma ^{12}-403178675 \gamma ^{11}
\nonumber \\ &-362536115 \gamma ^{10}+77856912 \gamma ^9+70236432 \gamma ^8+16701489 \gamma ^7 +16955505 \gamma ^6\nonumber \\ & -536235 \gamma ^5-536235 \gamma ^4+393120 \gamma ^3+393120 \gamma ^2+10395 \gamma +10395\Big],\nonumber \\ 
c_{2}(\gamma)&=-\frac{1}{2520 \gamma ^8 \left(\gamma ^2-1\right)^{9/2}}\Big[651264 \gamma ^{20}-7809042 \gamma ^{18}-23185512 \gamma ^{16}+169295016 \gamma ^{14}\nonumber \\  & -315460542 \gamma ^{12}+277369170 \gamma ^{10}-134264214 \gamma ^8+6510035 \gamma ^6-988015 \gamma ^4\nonumber \\  & +240905 \gamma ^2-18585\Big],\nonumber \\ 
c_{3}(\gamma)&=\frac{1}{360 (\gamma^{2} -1)^3 \gamma ^7 }\Big[6144 \gamma ^{16}-587336 \gamma ^{14}+4034092 \gamma ^{12}-417302 \gamma ^{10}-5560073 \gamma ^8\nonumber \\  & -142640 \gamma ^6+35710 \gamma ^4-8250 \gamma ^2+1575\Big],\nonumber \\ 
c_{4}(\gamma)&=-\frac{\gamma  \left(32768 \gamma ^8-90112 \gamma ^6+1564672 \gamma ^4-1872978 \gamma ^2-7817455\right)}{336 (\gamma^{2} -1)^2},\nonumber \\ 
c_{5}(\gamma)&=-\frac{1}{840 (\gamma^{2} -1)^5 \gamma ^7}\Big[491520 \gamma ^{22}-2482176 \gamma ^{20}+10655064 \gamma ^{18}-32742084 \gamma ^{16}\nonumber \\  & +17085516 \gamma ^{14}+61205662 \gamma ^{12}-59068870 \gamma ^{10}-5433687 \gamma ^8+1352120 \gamma ^6\nonumber \\ & -330890 \gamma ^4 +72450 \gamma ^2-11025 \Big],\nonumber \\ 
c_{6}(\gamma)&=-\frac{1}{120 \gamma ^8 \left(\gamma ^2-1\right)^{7/2}}\Big[24576 \gamma ^{18}+213480 \gamma ^{16}-1029342 \gamma ^{14}-1978290 \gamma ^{12}\nonumber \\  & +3752006 \gamma ^{10}+816595 \gamma ^8-55260 \gamma ^6+13690 \gamma ^4-3100 \gamma ^2+525\Big],\nonumber \\ 
c_{7}(\gamma)&=\frac{198856 \gamma ^{14}-689664 \gamma ^{12}-154716 \gamma ^{10}+666260 \gamma ^8-5091 \gamma ^6-1935 \gamma ^4+155 \gamma ^2-105}{12 \gamma ^8 \left(\gamma ^2-1\right)^{5/2}},\nonumber \\ 
c_{8}(\gamma)&=-\frac{1 }{42 \gamma ^6 \left(\gamma ^2-1\right)^{7/2}}\Big[49152 \gamma ^{18}-208896 \gamma ^{16}+1182464 \gamma ^{14}-3741239 \gamma ^{12}\nonumber \\  &  +3040161 \gamma ^{10}+1882567 \gamma ^8-2828161 \gamma ^6+49728 \gamma ^4-2268 \gamma ^2+1260\Big],\nonumber \\ 
c_{9}(\gamma)&=-\frac{\gamma  \left(525 \gamma ^8-450 \gamma ^6+17700 \gamma ^4-12598 \gamma ^2-5369\right)}{4 \left(\gamma ^2-1\right)^{7/2}},\nonumber \\ 
c_{10}(\gamma)&=-\frac{81920 \gamma ^6+189180 \gamma ^4-1240416 \gamma ^2-199207}{48 \left(\gamma ^2-1\right)^{5/2}},\nonumber \\ 
c_{11}(\gamma)&=-\frac{128 \gamma  \left(2 \gamma ^2-3\right) \left(8 \gamma ^6-6 \gamma ^4-51 \gamma ^2-8\right)}{(\gamma^{2} -1)^4 },\nonumber \\ 
c_{12}(\gamma)&=-\frac{\gamma  \left(2 \gamma ^2-3\right) \left(2273 \gamma ^6-1851 \gamma ^4-12957 \gamma ^2-2057\right)}{2 (\gamma^{2} -1)^4 },\nonumber \\ 
c_{13}(\gamma)&=-\frac{\gamma  \left(1575 \gamma ^6+1920 \gamma ^4-5177 \gamma ^2-1182\right)}{2 \left(\gamma ^2-1\right)^{5/2}},\nonumber \\ 
c_{14}(\gamma)&=-\frac{1249 \gamma ^6-1083 \gamma ^4-1053 \gamma ^2-9}{\left(\gamma ^2-1\right)^{5/2}},\nonumber \\ 
c_{15}(\gamma)&=-\frac{3 \left(225 \gamma ^6+600 \gamma ^5-315 \gamma ^4-1200 \gamma ^3+99 \gamma ^2+56 \gamma -9\right)}{\left(\gamma ^2-1\right)^{5/2}},\nonumber \\ 
c_{16}(\gamma)&=\frac{\gamma  \left(2100 \gamma ^6+1755 \gamma ^4-6422 \gamma ^2-1209\right)}{4 \left(\gamma ^2-1\right)^{5/2}},\nonumber \\ 
c_{17}(\gamma)&=\frac{9 \gamma  \left(2 \gamma ^2-3\right) \left(5 \gamma ^2-1\right)^2}{(\gamma^{2} -1)^3 },\nonumber \\ 
c_{18}(\gamma)&=\frac{1823 \gamma ^6-1221 \gamma ^4-13155 \gamma ^2-2039}{\left(\gamma ^2-1\right)^{5/2}},\nonumber \\ 
c_{19}(\gamma)&=\frac{\gamma  \left(2 \gamma ^2-3\right) \left(799 \gamma ^6-453 \gamma ^4-12003 \gamma ^2-2039\right)}
{(\gamma^{2} -1)^4 },\nonumber \\ 
c_{20}(\gamma)&=\frac{768 \left(8 \gamma ^6+14 \gamma ^4-116 \gamma ^2-19\right)}{\left(\gamma ^2-1\right)^{5/2}},\nonumber \\ 
c_{21}(\gamma)&=-\frac{3 \gamma ^2 \left(2 \gamma ^2-3\right) \left(175 \gamma ^6-355 \gamma ^4+185 \gamma ^2-37\right)}{4 (\gamma^{2} -1)^4 },\nonumber \\ 
c_{22}(\gamma)&=-\frac{4 \left(1823 \gamma ^6+6459 \gamma ^4-38115 \gamma ^2-6263\right)}{\left(\gamma ^2-1\right)^{5/2}},\nonumber \\ 
c_{23}(\gamma)&=-\frac{3871 \gamma ^6-2817 \gamma ^4-3747 \gamma ^2-75}{2 \left(\gamma ^2-1\right)^{5/2}},\nonumber \\ 
c_{24}(\gamma)&=-\frac{3 \gamma  \left(175 \gamma ^6-1255 \gamma ^4+1985 \gamma ^2-121\right)}{4 \left(\gamma ^2-1\right)^{5/2}},\nonumber \\ 
c_{25}(\gamma)&=\frac{3421 \gamma ^6-2067 \gamma ^4-5865 \gamma ^2+111}{2 \left(\gamma ^2-1\right)^{5/2}},\nonumber \\ 
c_{26}(\gamma)&=\frac{48 \gamma  \left(75 \gamma ^4-150 \gamma ^2+7\right)}{\left(\gamma ^2-1\right)^{5/2}},\nonumber \\ 
c_{27}(\gamma)&=-\frac{2 \left(4321 \gamma ^6+11973 \gamma ^4-75933 \gamma ^2-12553\right)}{\left(\gamma ^2-1\right)^{5/2}},\nonumber \\ 
c_{28}(\gamma)&=-\frac{64 \left(160 \gamma ^{12}-600 \gamma ^{10}-84 \gamma ^8+2058 \gamma ^6-1665 \gamma ^4-72 \gamma ^2+32\right)}{3 \left(\gamma ^2-1\right)^{11/2}},\nonumber \\ 
c_{29}(\gamma)&=-\frac{\gamma  \left(2 \gamma ^2-3\right) \left(1823 \gamma ^6-1221 \gamma ^4-13155 \gamma ^2-2039\right)}{(\gamma^{2} -1)^4 },\nonumber \\ 
c_{30}(\gamma)&=\frac{3150 \gamma ^7+1846 \gamma ^6-5775 \gamma ^5+198 \gamma ^4+5488 \gamma ^3-7518 \gamma ^2-1935 \gamma +258}{8 \left(\gamma ^2-1\right)^{5/2}},\nonumber \\ 
c_{31}(\gamma)&=\frac{1050 \gamma ^6+1696 \gamma ^5+389 \gamma ^4-1691 \gamma ^3-2241 \gamma ^2-1041 \gamma -114}{8 (\gamma -1) (\gamma ^2-1)^{3/2}},  \end{align}
where the $G(a_{1},......, a_{n};y)$ are the multiple polylogarithms defined as $ G(a_{1},\ldots,a_{n};y) = \int_{0}^{y} \frac{d t}{t-a_{1}} G(a_{2},\ldots,a_{n};t)$ 
  and  $y=1 -\gamma +\sqrt{\gamma^2-1}$. 

In Eq.  (\ref{2pmchi}),  we used 
  \begin{eqnarray}
&&\text{Li}_2(z) \equiv \int_z^0 dt \, \frac{\log(1-t)}{t}\,,   \, \nonumber \\
&& \mathrm{K}(z) \equiv 
 \int_0^1 \frac{dt}{\sqrt{\left(1-t^2\right)\left(1-z t^2\right)}}\,,\nonumber \\
&&\mathrm{E}(z) \equiv  \int_0^1 dt\, \frac{\sqrt{1-z t^2}}{\sqrt{1-t^2}}\,,
  \end{eqnarray}
for the dilogarithm, and complete elliptic integral of the first and second kind, respectively.

%%%%%%%%%%%%%%%%%%%%%%%

\section{The quantities  \texorpdfstring{ $J_j $}{} (j=2, 3, 4), and  \texorpdfstring{ $E_k$}{} (k=3,4)  in the equation (\ref{J234})} \label{Jii}

\begin{eqnarray}
J_2&=& \frac{4(2\gamma^2-1)\sqrt{\gamma^2-1}}{\Gamma^2}\Big[ -\frac{8}{3}+ \frac{ \gamma^2}{\gamma^2-1}+ \frac{\gamma (2 \gamma^2-3)}{(\gamma^2-1)^{3/2}} {\rm arctanh}(\sqrt{1- \frac1{\gamma^2}}) \Big]\,, \nonumber \\ 
%J_3&=&\frac{\widehat J_3}{\Gamma^3} -\frac{\nu  E_3}{\Gamma}, \nonumber \\ 
J_3&=&\frac{\pi}{\Gamma^3}\Bigg[
\frac{  \left(-141+386 \gamma -525 \gamma ^2-683 \gamma ^3+1377 \gamma ^4+240 \gamma ^5-711 \gamma ^6+105 \gamma ^7\right)}{24 \left(-1+\gamma
^2\right)}\nonumber \\ &-&\frac{  \gamma  \left(-3+2 \gamma ^2\right) \left(-24+19 \gamma +144 \gamma ^2-70 \gamma ^3-120 \gamma ^4+35 \gamma ^5\right) \text{arccosh}[\gamma
]}{8 \left(-1+\gamma ^2\right)^{3/2}}\nonumber \\ &+&\frac{1}{4}  \left(-62+155 \gamma +16 \gamma ^2-70 \gamma ^3-90 \gamma ^4+35 \gamma ^5\right) \text{Log}\left(\frac{1+\gamma
}{2}\right)\Bigg] 
-\frac{\nu  E_3}{\Gamma}, \nonumber \\ %J_4&=&\frac{\widehat J_4}{\Gamma^4} -\frac{\nu E_4}{ \Gamma}, \nonumber \\
J_4&=&\frac{1}{\Gamma^4}\Bigg\{ \frac{176 }{5}\sqrt{\gamma^2-1}+\frac{8144}{105}(\gamma^2-1)^{3/2}+\frac{448 }{5}(\gamma^2-1)^2
-\frac{93664}{1575}(\gamma^2-1)^{5/2}\nonumber \\& +&\frac{1184}{21} (\gamma^2-1)^3- \frac{4955072}{121275} (\gamma^2-1)^{7/2}-\frac{13736}{315} (\gamma^2-1)^4+\nu \Big[ -\frac{208}{15} (\gamma^2-1)^{3/2}\nonumber \\ &+& \frac{988}{63} (\gamma^2-1)^{5/2}-\frac{13312}{525} (\gamma^2-1)^3+\frac{5458}{1575} (\gamma^2-1)^{7/2} - \frac{208}{225} (\gamma^2-1)^4\Big]\Bigg\} -\frac{\nu E_4}{ \Gamma}, \nonumber \\
E_3&=&\frac{\pi (\gamma^2-1)^{3/2}}{\Gamma^4} \Big[\frac{ (1151-3336\gamma+3148\gamma^2
-912\gamma^3+339\gamma^4-552\gamma^5+210\gamma^6)}{48(\gamma^2-1)^{3/2}}\nonumber \\ &-& \frac{76\gamma -150\gamma^2+60\gamma^3+35\gamma^4-5}{8\sqrt{\gamma^2-1}}\ln\frac{\gamma+1}{2}
- \frac{\gamma (3-2\gamma^2) (11-30\gamma^2+35\gamma^4)}{16(\gamma^2-1)^2} {\rm arccosh}(\gamma)\Big], \nonumber \\ 
E_4&=& \frac{1}{\Gamma^5}\Big[\frac{1568}{45} (\gamma^2-1)^{3/2}+\frac{18608}{525} (\gamma^2-1)^{5/2}+\frac{3136}{45} (\gamma^2-1)^3+\frac{220348}{11025} (\gamma^2-1)^{7/2}\nonumber \\ &+&\frac{1216
  }{105} (\gamma^2-1)^4-\frac{151854}{13475} (\gamma^2-1)^{9/2}+\frac{117248}{1575} (\gamma^2-1)^5-\frac{405087523
   }{9909900}(\gamma^2-1)^{11/2}\nonumber \\ & - & \frac{224512}{3465} (\gamma^2-1)^6\Big]+\nu \Big[
  -\frac{352}{45} (\gamma^2-1)^{5/2}+\frac{1736
   }{225}(\gamma^2-1)^{7/2}-\frac{704}{45} (\gamma^2-1)^4\nonumber \\ &+& \frac{8068}{4725} (\gamma^2-1)^{9/2}-\frac{1808}{225} (\gamma^2-1)^5+\frac{1967239
   }{103950} (\gamma^2-1)^{11/2}-\frac{47176 }{4725}(\gamma^2-1)^6\Big]. 
\end{eqnarray}
Please note that $J_4$ and $E_4$ are just approximate results. 

%%%%%%%%%%%%%%%%%%%

\section{Calculations of the scattering angle for  EOB system}

By using Eqs.~\eqref{expansionB} and~\eqref{effR0}, up to 5PM order, we find $\mathcal{R}_0(R,\mathcal{E}_0,J_0)$ can be expressed as
\begin{eqnarray}\label{eobaction}
\mathcal{R}_0(r,\mathcal{E}_0,J_0)&=&R_{00}-\frac{J_0^2}{r^2}+\frac{R_{11} G}{r}+\frac{R_{22} G^2}{r^2}-\frac{R_{31} J_0^2 G -R_{33} G^3 }{r^3}-\frac{R_{42}J_0^2 G^2-R_{44} G^4}{r^4}\nonumber \\&-&\frac{R_{53}J_0^2 G^3-R_{55} G^5}{r^5}-\frac{R_{64}J_0^2 G^4}{r^6}-\frac{R_{75}J_0^2 G^5}{r^7}\;,
\end{eqnarray}
with
\begin{eqnarray}\label{Rii}
R_{00}&=&\mathcal{E}_0^2-m_0^2\;,\nonumber \\
R_{11}&=&2 M_0\left[2 \mathcal{E}_0^2-m_0^2\right] \;,
\nonumber \\
R_{22}&=&M_0^2\left[(a_2-4) m_0 ^2+2 \mathcal{E}_0^2\left(6-a_2\right)\right]\;,\nonumber \\
R_{31}&=&2 M_0  \:,\nonumber\\
R_{33}&=&M_0^3\left[(4 a_2+a_3-8) m_0^2+2 \mathcal{E}_0^2\left(16-6a_2-a_3\right)\right]\;,\nonumber \\
R_{42}&=&(4-a_2 )  M_0^2\;,\nonumber \\
R_{44}&=&-
M_0^4 \Big\{(16-12 a_2+a_2^2-4 a_3 -a_4) m_0^2- \mathcal{E}_0^2  \Big[80+ 3 a_2^2-12 a_3-2 a_4 -48 a_2
 \Big]\Big\}\;,\nonumber \\
R_{53}&=&(8-4 a_2-a_3) M_0^3\;,\nonumber \\
R_{55}&=& \Big(\mathcal{E}_0^2 \left(20
   a_{2}^2+a_{2} (5 a_{3}-136)-2 (22
   a_{3}+6
   a_{4}+a_{5}-80)\right)\nonumber \\ 
   &+&m_{0}^2
   \left(-6 a_{2}^2-2 a_{2}
   (a_{3}-16)+12 a_{3}+4
   a_{4}+a_{5}-32\right)\Big)M_{0}^5, \nonumber \\ 
   R_{64}&=&(16-12 a_2+a_2^2-4 a_3 -a_4)  M_0^4\;,\nonumber \\
R_{75}&=& \Big(6 a_{2}^2+2
   a_{2} (a_{3}-16)-12 a_{3}-4
  a_{4}-a_{5}+32\Big) M_{0}^5,
   \end{eqnarray}
where  we have taken $a_1=-2$ by requiring that the effective metric at the 1PM order coincides with the Schwarzschild metric for the sake of concise expressions.

For the EOB system, the  scattering angle is defined as
\begin{eqnarray}\label{chiiC}
\chi^{\text{eff}}&=&-\pi-2  \int_{r_{\rm min}}^{\infty}\frac{\partial\sqrt{ \mathcal{R}_0 (r,\mathcal{E}_0,J_0)}}{\partial J_0} d r.
\end{eqnarray}
where $r_\text{min}$ is the minimum distance  determined by setting the vanishing of $\mathcal{R}_0 (r,\mathcal{E}_0,J_0)$.
Using Eqs. (\ref{eobaction})  and  (\ref{chiiC})  and  working out the integration up to the 5PM order, we have
\begin{eqnarray}\label{chiiA}
\chi^{\text{eff}}&=&\frac{R_{11}}{\sqrt{R_{00}}}\frac{G}{J_0}+
\frac{ \pi \big[2 R_{22} -  R_{11} R_{31} + R_{00}(\frac{3}{4}  R_{31}^2-R_{42})\big]}{4}
 \Big(\frac{G}{J_0}\Big)^2\nonumber \\
  &+&\frac{1}{12 R_{00}^{3/2}} \Big\{-R_{11}^3 - 6 R_{00} R_{11}^2 R_{31} + 12 R_{00} R_{11} \Big[R_{22} \nonumber \\ &+& 2 R_{00} (R_{31}^2 - R_{42})\Big ] -
   8 R_{00}^2 \Big[3 R_{22} R_{31} - 3 R_{33} \nonumber \\
 &+ &2 R_{00} (R_{31}^3 - 2 R_{31} R_{42} + R_{53})\Big]\Big\}\Big(\frac{G}{J_0}\Big)^3\nonumber\\
&+& \frac{3\pi }{1024}\Big\{128 R_{22}^2+48 R_{11}^2 \left(5 R_{31}^2-4 R_{42}\right)
\nonumber\\
&-&96 R_{22} \left(4 R_{11} R_{31}-5 R_{00}
R_{31}^2+4 R_{00} R_{42}\right)
\nonumber\\
&+&16 R_{11} \left(16 R_{33}+R_{00}\left(-35 R_{31}^3+60 R_{31} R_{42}-24
R_{53}\right)\right)
\nonumber\\
&+&R_{00} \Big[-384 R_{31} R_{33}+256 R_{44}+3 R_{00} \Big(105 R_{31}^4
\nonumber\\
&-&280 R_{31}^2
R_{42}+80 R_{42}^2+160 R_{31} R_{53}-64 R_{64}\Big)\Big]\Big\} \Big(\frac{G}{J_0}\Big)^4\nonumber \\ 
&+&\frac{1}{240 R_{00}^{5/2}}\Big[-64 R_{00}^3 \big(5 R_{22} \left(4 R_{00} \left(2
   R_{31}^3-3 R_{31} R_{42}+R_{53}\right)-3
   R_{33}\right)\nonumber \\ 
&+&2 R_{00} \big(4 R_{00} \big(3 R_{31}^5-10
   R_{31}^3 R_{42}+6 R_{31}^2 R_{53}+6 R_{31}
   R_{42}^2-3 R_{31} R_{64}\nonumber \\ 
&-&3 R_{42}
   R_{53}+R_{75}\big)-5 \left(3 R_{31}^2 R_{33}-2
   R_{31} R_{44}-2 R_{33} R_{42}+R_{55}\right)\big)+15
   R_{22}^2 R_{31}\big)\nonumber \\ 
&-&240 R_{00}^2 R_{11}^2 \left(4
   R_{00} \left(2 R_{31}^3-3 R_{31}
   R_{42}+R_{53}\right)+2 R_{22} R_{31}-R_{33}\right)\nonumber \\ 
&+&80
   R_{00}^2 R_{11} \big(12 R_{00} R_{22} \left(3 R_{31}^2-2
   R_{42}\right)+4 R_{00} \big(2 R_{00} \big(5 R_{31}^4-12
   R_{31}^2 R_{42}\nonumber \\ 
&+& 6 R_{31} R_{53}+3 R_{42}^2-2
   R_{64}\big)-6 R_{31} R_{33}+3 R_{44}\big)+3
   R_{22}^2\big)+20 R_{00} R_{11}^4 R_{31}\nonumber \\ 
&+&40 R_{00}
   R_{11}^3 \left(6 R_{00} R_{31}^2-4 R_{00}
   R_{42}-R_{22}\right)+3 R_{11}^5\Big]\Big(\frac{G}{J_0}\Big)^5.
 \end{eqnarray}

%\end{appendices}

\bibliography{mybib}

\end{document}